# Synergistic effects of MoDTC and ZDTP on frictional behaviour of tribofilms at the nanometer scale

S. Bec[1*], A. Tonck[1], J.M. Georges[1] and G.W. Roper[2]

[1]*Laboratoire de Tribologie et Dynamique des Systèmes, UMR CNRS 5513, Ecole Centrale de Lyon, 36 av. Guy de Collongue, 69134 Ecully Cedex, France.*

[2]*Lubricants Technology Dept., Shell Global Solutions, Shell Research and Technology Centre, Thornton, P. O. Box 1, Chester CH1 3SH, UK.*

[*]*To whom correspondence should be addressed*

**Abstract**
The layered structure and the rheological properties of anti-wear films generated in a rolling/sliding contact from lubricants containing zinc dialkyldithiophosphate (ZDTP) and/or molybdenum dialkyldithiocarbamate (MoDTC) additives have been studied by dynamic nanoindentation experiments coupled with a simple modelling of the stiffness measurements. Local nano-friction experiments were conducted with the same device in order to determine the evolution of the friction coefficient as a function of the applied pressure for the different lubricant formulations. For the MoDTC film, the applied pressure in the friction test remains low (<0.5 GPa) and the apparent friction coefficient is high (µ>0.4). For the tribofilms containing MoDTC together with ZDTP, which permits the applied pressure to increase up to a few GPa through some accommodation process, a very low friction domain appears (0.01<µ<0.05), located a few nanometers below the surface of the tribofilm. This low friction coefficient is attributed to the presence of $MoS_2$ planes sliding over each other in a favourable configuration obtained when the pressure is sufficiently high, which is made possible by the presence of ZDTP.

## 1. Introduction

In addition to zinc dialkyldithiophosphate (ZDTP) additives, extensively used for their exceptional antioxidant and anti-wear properties under boundary conditions in automotive engines, lubricating oils contain several additives, among which there are detergent and dispersant additives whose main role is to keep oil insoluble contaminants and degradation products in suspension, at elevated temperature for the detergent additives, and at low temperatures for the dispersant ones. Organo molybdenum compounds such as molybdenum dithiocarbamate (MoDTC) are also used as friction modifiers for energy saving. However, when used together in formulated oils, additives interact in various ways resulting either in synergies or in adverse effects affecting the oil performance regarding anti-wear and friction behaviour, and modifying the characteristics of the protective surface films generated during friction (tribofilms). A lot of investigations have been conducted to evaluate the performances of additive mixtures and to determine the composition of associated tribofilms. Several factors were identified as playing a role: additive structure [1, 2], additives concentration [3-6], base oil nature [7, 8], …, or combinations of these parameters. A detailed review on published information on that topic was written by Willermet [9]. Non-chemical parameters such as characteristics of the solid antagonists (hardness, roughness) or test conditions (load, temperature, sliding speed) [3, 10] also might influence the additive interactions.

Among this variety of additive interactions, we will focus on that between ZDTP and MoDTC, extensively studied through chemical investigations. All published works agree upon the fact that friction and anti-wear performances of oils are improved when ZDTP and MoDTC are used together. The formation of molybdenum disulphide ($MoS_2$) on the rubbing surfaces has been evidenced by several authors [11, 12]. Using UHV friction tests, coupled with high-resolution TEM observation of wear debris and spectroscopic studies, Grossiord et al. has given evidence for the mechanism of single $MoS_2$ sheet lubrication [13].

The aim of this paper is to enlarge the knowledge of the local mechanical and frictional properties of anti-wear tribofilms to those of films obtained from lubricants containing different additives (ZDTP, MoDTC, detergent/dispersant) or mixtures of additives, in order to explore the ZDTP/MoDTC synergy on a mechanical point of view. The only published results on that topic are the recent papers from Ye et al. who performed AFM observations and nanoindentation measurements on ZDTP and ZDTP + MoDTC tribofilms [14, 15].

In the present study, nanoindentation tests with continuous stiffness measurements were performed on unwashed and solvent-washed tribofilms to determine their mechanical properties. The frictional behaviour of the tribofilms was investigated through local nanofriction experiments, conducted with the same device. The evolution of the friction coefficient as a function of the applied pressure for the different lubricant formulations leading to different tribofilms has been determined.

## 2. Preliminary results obtained on ZDTP anti-wear tribofilms

The structure and the rheological properties of anti-wear films from a zinc dialkyldithiophosphate (ZDTP) solution generated in a rolling/sliding contact, simulating engine valve train conditions, have been studied in detail with analytical and surface force tools and the results have been published by the authors in a previous paper [16]. As preamble to the present paper, only the main points are summarised here.

The ZDTP solution was a commercial secondary alkyl ZDTP additive at 0.1% weight of phosphorus in a highly refined base oil. The ZDTP anti-wear films have a complex structure that has been determined by extensive use of surface analytical techniques. It has been shown that the ZDTP films consisted of at least three non-homogeneous layers: on the steel surface, there is a sulphide/oxide layer, which is almost completely covered by a protective phosphate layer, with the addition of a viscous overlayer of ZDTP degradation precipitates (alkyl phosphate precipitates). This latter layer was removed when the film was washed with an alkane solvent. Therefore, the properties of the ZDTP films have been studied both before and after solvent washing with n-heptane. First, sphere/plane squeeze experiments were performed with a surface force apparatus (SFA) on unwashed films, showing that the overlayer of alkyl phosphate precipitates was heterogeneous and discontinuous, with a thickness of about 900 nm. Second, the mechanical properties were obtained from nanoindentation experiments, performed after replacing the sphere by a diamond tip, and coupled with in-situ topographic imaging procedures to measure the contact area. From the indentation experiments, the properties of the films were determined from normal stiffness measurements and through the application of a rheological film model. On the unwashed specimens, the viscous layer of alkyl phosphate precipitates was detected by the indentation tests. It is a very soft layer, mobile under the diamond tip, with a thickness of a few hundred of nanometers, which was in good agreement with that of sphere/plane experiments. It was also shown that indentation experiments removed this overlayer in the proximity of the tip, probably through a shear flow mechanism. This procedure can be compared to a soft "mechanical" sweep and the mechanical properties of ZDTP tribofilms after such a cleaning were found to be similar to those of solvent washed specimens. The solvent washed tribofilms, comprising sulphide and phosphate layers, exhibited an elastoplastic behaviour and, during the loading stage of the indentation, the hardness and the Young's modulus of the phosphate layer increased from their initial values of about 2 GPa for the hardness and between 30 and 40 GPa for the Young's modulus. In particular, the initial hardness of the polyphosphate layer at the beginning of the indentation tests was close to the mean applied pressure during the films generation. This suggested that the layer accommodated the contact pressure in the tribotest or during the loading stage of the indentation, and could thus be regarded as a final and local pressure sensor. The characteristics of the full ZDTP films ensure gradual changes in mechanical properties between the substrate, bonding layers and outer layers with the viscous overlayer serving as the tribofilm's precursor. The properties of these layered films can thus adapt to a wide range of imposed conditions and provide appropriate level of resistance to contact between the metal surfaces. As the severity of loading increases, so too do the resistive forces within the film. This ensures that the shear plane remains located inside the ZDTP protective film, which explains the exceptional efficiency of ZDTP films as anti-wear films.

**3. Experimental**
*3.1. Tribofilms*
The tribofilms were generated at Shell Research and Technology Centre, Thornton, U.K., with a reciprocating Amsler machine [17] designed to simulate the contact conditions of the cam/follower system in an internal combustion engine valve train. A flat block specimen (8 mm x 8 mm size, 4 mm thick) has a reciprocating motion in loaded contact with a rotating disc. The block and the disc were made in through-hardened EN31 steel. Special care was taken with the roughness of the blocks which were polished until the average roughness was Ra = 0.01 µm. The movement of the block was driven by a crank linked to the motion of the disc axis through a gearbox. The block motion was approximately sinusoidal and at the same frequency as the disc rotation. Load was applied to the contact by a spring arrangement, acting through a roller bearing. The surface in contact with the loading bearing (rear surface of the reciprocating element) was curved to permit self-alignment between the block and the disc. The films were generated at a normal load of 400 N (mean contact pressure of 0.36 GPa), speed of 600 rev/min., block temperature of approximately 100°C for 5 hours. The lubricants consisted of a highly refined base oil with different commercial additives (details of the oil formulation are not relevant to the present work):
- MoDTC solution,
- ZDTP + MoDTC solution,
- ZDTP + MoDTC + detergent/dispersant solution ("full formulation").

The rubbing area on the polished block was typically 5 mm long in the sliding direction. Previous analyses have shown that the composition in the centre of the wear track was reasonably uniform, while the composition within 1 mm of the ends of the wear track could vary significantly. The mechanical measurements on the films with the Surface Force Apparatus have been performed in the central area of the wear track. An additional unworn and polished block was used to obtain reference values for the EN31 steel substrate.

To preserve the film structures, the blocks were stored in the base oil (containing predominantly paraffinic hydrocarbons, with very low concentration of polar compounds) immediately after production of the films in the reciprocating Amsler tests and they were immersed again, when not in use.

*3.2. Surface Force Apparatus*
The Ecole Centrale de Lyon Surface Force Apparatus (SFA) used in these experiments has been described in previous publications [18, 19]. The general principle is that a macroscopic spherical body or a diamond tip can be moved toward and away from a planar one (the ZDTP specimen) using the expansion and the vibration of a piezoelectric crystal, along the three directions, Ox, Oy (parallel to the plane surface) and Oz (normal to the plane surface). The plane specimen is supported by double cantilever sensors, measuring quasi-static normal and tangential forces (respectively $F_z$ and $F_x$). Each of these is equipped with a capacitive sensor. The sensor's high resolution allows a very low compliance to be used for the force measurement (up to $2 \times 10^{-6}$ m/N). Three capacitive sensors were designed to measure relative displacements in the three directions between the supports of the two solids, with a resolution of 0.01 nm in each direction. Each sensor capacitance was determined by incorporating it in an LC oscillator operating in the range 5 - 12 MHz [20].

*3.3. Tests methodology*

All the experiments were conducted at room temperature. Preliminary results obtained on anti-wear films from a ZDTP solutions have shown that n-heptane washing damages the film [16]. That is why the blocks were tested first as obtained from the Amsler friction test, without any cleaning and second after washing with n-heptane. The unwashed specimens were mounted on the SFA as taken from the storage base oil. Excess of base oil was simply removed by placing the side of the specimen on absorbing paper, which allowed the surface to be always preserved by an oil film (thickness > 10 µm).

Nanoindentation tests

The aim of these tests was to determine the elastoplastic properties of the tribofilms (hardness and Young's modulus) and their "mechanical" structure (number of layers and estimation of the thickness of each layer that constitutes the film). The method used to perform nanoindentation experiment with the SFA has already been published in detail [21]. Specific procedures have been developed for the characterisation of ZDTP tribofilms and have been described in previous papers [16, 22]. In this study, the determination of the near surface mechanical properties (first nanometers) was obtained through a specific tip shape calibration, performed on a gold film deposited by magnetron sputtering onto a silicon substrate. This film was very smooth (peak to valley roughness around 1 nm, measured on a scan length of 1 µm) and its hardness was constant versus depth from the surface and until the penetration depth equals the gold film's thickness [21].

For the nanoindentation experiments, a trigonal diamond tip with an angle of 115.12° between edges (Berkovitch type) was used. The indentation tests were performed in controlled displacement mode. The standard set-up included the continuous quasi-static measurements of the resulting normal force $F_z$ versus the normal displacement Z, at a slow penetration speed, generally 0.1 to 0.5 nm/s. It also included the simultaneous measurements of the rheological behaviour (dissipative and conservative or elastic contributions) of the tested surface, thanks to simultaneous small sinusoidal motions at a frequency of 37 Hz, with an amplitude of about 0.2 nm RMS. Furthermore, using the Z feedback in the constant force mode and the tangential displacement of the indenter, the surface topography was imaged before and after the indentation test, with the same diamond tip. This was made practically possible because of the partial elastic recovery during the unloading cycle and hence the geometry of tip and indent were different which was necessary to permit resolution of the indent. For this scanning procedure, a constant normal load of 0.5 µN was typically used. Such in-situ imaging procedure enables the operator to choose precisely the location of the indentation test on the surface and, after the test, to quantify the plastic pile-up around the indent and thus to measure the actual contact area.

Rheological film model

The elastic properties of the films were very difficult to extract from the indentation tests because of both the influence of the substrate and of the film structure itself. They were obtained through the stiffness measurements, which are global (film+substrate) measurements. To extract the properties of each layer of the film, a simple model has been developed, and its main features are described as follows. The experimental stiffness versus normal displacement curve was identified with the elastic response of a structure composed of one or two homogeneous elastic layers on a substrate (semi-infinite elastic half space) indented by a rigid cylindrical punch of radius a. For such a system, modelled by two springs connected in series [23], the calculated global stiffness ($K_z$) depends on the reduced Young's modulus of the substrate ($E_s^*$ with $E_s^*=E_s/(1-v_s^2)$), measured on an unworn steel block, over the contact radius (a) and depends also on four unknown parameters which are the reduced Young's modulus ($E_f^*$, $E_f^*=E_f/(1-v_f^2)$) and the thickness (t) of each layer. For each test, their values were adjusted to obtain a good fit between the measured stiffness curve and the calculated one. This procedure provided the structure (one or two layers), the thickness and the reduced Young's modulus of each layer that constituted the tribofilms. Details are given in a previous paper [16]. Following this model, the global stiffness of a single layer system is given by:

$$\frac{1}{K_z} = \frac{1}{1+2t/\pi a}\left(\frac{t}{\pi a^2 E_f^*} + \frac{1}{2aE_s^*}\right) \quad (1)$$

This simple model describes perfectly the behaviour of model systems such as gold layers on a silicon substrate [21]. In the case of tribofilms, deviations may be observed at a critical pressure or at a critical depth from which the experimentally measured stiffness may be found to exceed significantly the theoretical one. This is interpreted as a change in the surface properties due to the applied pressure and appears to be related to a measured hardness increase. Indeed, as the applied pressure can reach values much larger than the initial hardness value of the surface, the resulting plastic flow may induce a small volume reduction and molecular rearrangements which could be sufficient to induce a noticeable change in the mechanical properties. From a threshold pressure value, $H_0$, the stiffness curve was then influenced both by the substrate's elasticity and by the change in mechanical properties. This pressure dependence can be introduced in the model by writing that in the deformed volume of material, when $H>H_0$ (i.e. when the film accommodates the applied pressure through hardness increase), the film modulus $E_f^*$ is proportional to the hardness (the ratio $E_f^*/H$ remains constant). It gives the following equation:

$$E_f^* = E_{f0}^* \frac{H}{H_0} \quad (2)$$

$E_{f0}^*$ is the reduced Young's modulus value, when the applied pressure is equal to or lower than the threshold pressure $H_0$. When necessary, by introducing this effect in our modelling and by adjusting the value of the threshold pressure, we were able to fit correctly the whole stiffness curve. An example of such fit is given on figure 1. The evolution of the film modulus $E_f^*$ versus plastic depth can also be extracted from equation 1 using the experimentally measured global (film+substrate) stiffness values $K_z$ and the film's thickness, t, independently of equation 2. This permits a check on whether it is proportional to the hardness as assumed in equation 2. In the example shown figure 2, the calculated Young's modulus of the film (from equation 1 with a film thickness t=25 nm) is found to be proportional to the measured hardness with a mean ratio $E_f^*/H=16.5$, in good agreement with the ratio $E_{f0}^*/H_0=17/1.05=16.2$ obtained from the stiffness fit.

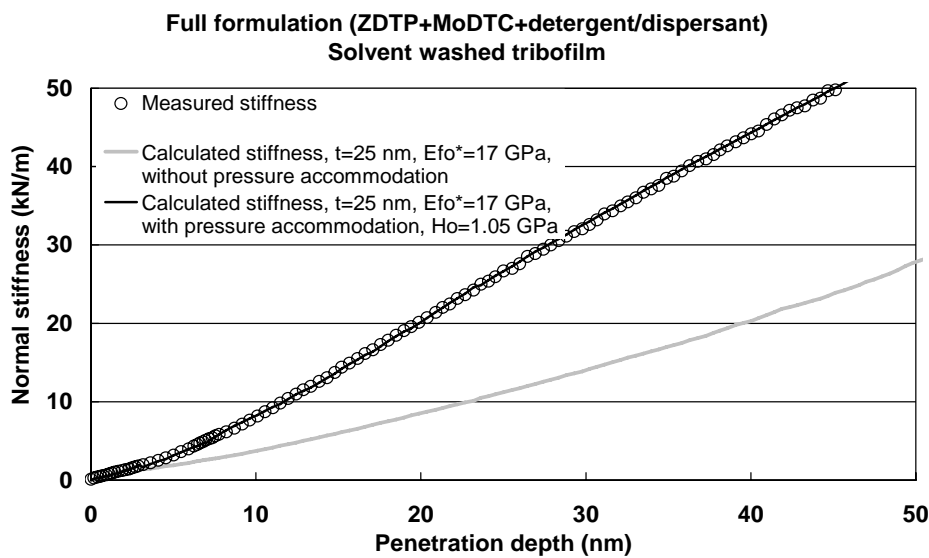

*Figure 1*: Example of application of the rheological film model: measured and calculated global stiffness for a tribofilm obtained from the full formulation (MoDTC + ZDTP + detergent/dispersant). A good fit between the measured and the calculated values is obtained with a single layer system (thickness t=25 nm and reduced Young's modulus $E_{f0}^*$=17 GPa) and a pressure accommodation effect from a threshold pressure $H_0$=1.05 GPa.

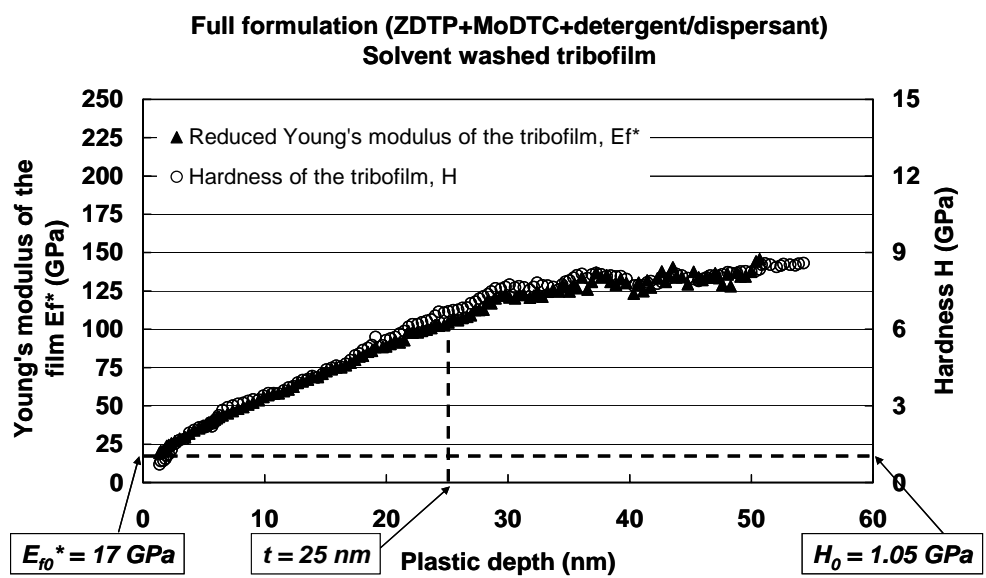

*Figure 2*: Example of evolution of film's reduced Young's modulus and hardness versus plastic depth, for a tribofilm obtained from the "full formulation" (MoDTC + ZDTP + detergent/dispersant). The Young's modulus of the film is calculated using equation 1 with the measured stiffness values and using only the film's thickness determined from the fit shown figure 1 (t = 25 nm).

Nanofriction experiments
Nanofriction experiments were conducted on the blocks by moving the diamond tip along Ox direction (parallel to the surface) at low speed (2 to 5 nm/s) along a distance of 0.5 µm. The objective of these tests was to determine how the friction coefficient varies as a function of the applied pressure. The tests were conducted at monitored increasing depth. During the tests, the normal, $F_z$, and the tangential, $F_x$, forces were recorded, which allowed us to calculate the apparent friction coefficient $\mu=F_x/F_z$ (see example figure 3).

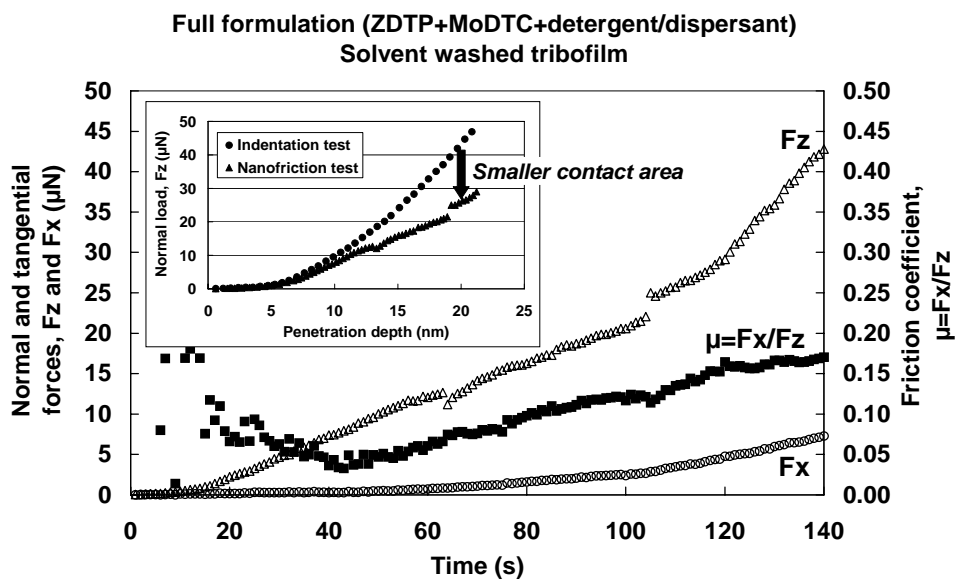

*Figure 3:* Procedure used for the nanofriction tests. The diamond tip is oriented edge first and the nanofriction tests are conducted at monitored increasing depth. During the test, the normal (Fz) and tangential (Fx) forces are recorded. The friction coefficient $\mu=Fx/Fz$ is calculated.

Large pile-up was observed in the case of nanofriction with the diamond tip oriented face first, which may induce large uncertainty in the calculation of the contact area. That is why the nanofriction tests were conducted edge first. In these conditions, the estimation of the applied pressure at a given depth was obtained using low load nanoindentation tests, made in the near proximity of the nanofriction tests. Assuming that, at a given depth, the hardness of the tribofilm should be the same for the friction test and for the near indentation test, the contact area, and then the applied pressure, were obtained from the difference between the normal force measured for the two tests at the same depth (see insert on figure 3). Using the in-situ imaging procedure, figure 4 shows an example of an image of the surface of a tribofilm after such a nanofriction test.

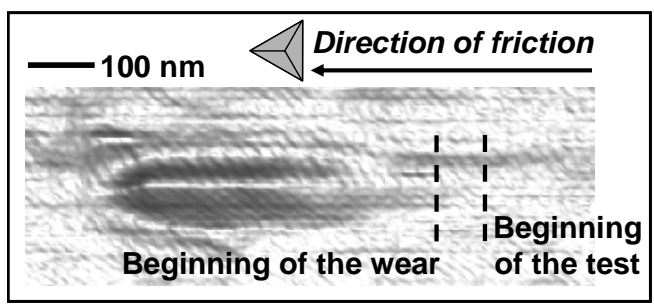

*Figure 4:* Typical image of the surface of a tribofilm after a nanofriction experiment. The image is obtained with the in-situ imaging procedure.

### 4. Results
The first part presents the mechanical properties of the different tribofilms, determined from the nanoindentation experiments. Their structure, one or two layers, and their thickness were deduced from the use of our rheological film model.

The results concerning the frictional behaviour of the tribofilms are given in a second part.

*4.1. Structure and mechanical properties of the tribofilms*

MoDTC tribofilms
The tribofilm obtained from base oil + MoDTC has been tested without washing and after washing with n-heptane. Even on the solvent washed block, it was not possible to make any local topographic image nor line scanning preliminary to the indentations tests, revealing that the film was very soft and was easily damaged by the diamond tip. Representative hardness curves obtained on the MoDTC tribofilms are shown on figure 5.

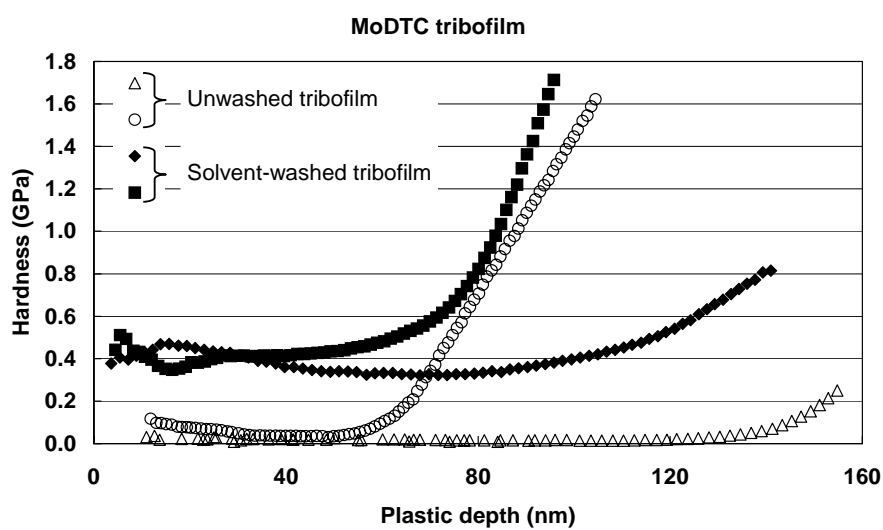

*Figure 5: Typical hardness curves obtained on the MoDTC tribofilms. Open symbols correspond to hardness curves obtained on the unwashed film. Black symbols correspond to hardness curves obtained on the solvent-washed film.*

Very low mechanical properties were measured on the unwashed MoDTC tribofilm. The surface hardness ranged from 0.02 to 0.1 GPa indicating the presence of a very soft overlayer covering the tribofilm.
After washing with n-heptane, the indentation tests showed that this overlayer has been removed by the washing procedure. The remaining tribofilm was a soft homogeneous layer, whose hardness was typically in the range 0.4 - 0.5 GPa at the beginning of the tests. Adhesion to the diamond tip was detected at the end of the unloading part of the tests. The film thickness and the structure (number of layers) have been obtained from the stiffness measurements performed during the experiments using the rheological film model.
The film appeared to be homogeneous in its thickness, and for most of the tests, its elastic behaviour corresponded to that of a single layer, with constant properties versus depth. The thickness of the film was found to be between 30 and 75 nm. The reduced Young's modulus was typically equal to 7 – 8 GPa.

ZDTP + MoDTC tribofilms
From optical observation, the ZDTP + MoDTC unwashed film was very thin. This was confirmed by the indentation tests. Prior to any contact, a very soft layer, 60 to 120 nm thick, was detected at the surface of the unwashed film.
Indentation tests conducted after scanning or imaging the surface of the unwashed film ("mechanical sweep") showed that the film was spatially heterogeneous. Its thickness and its mechanical properties varied depending on the test location:
- In some places, only a very thin layer (a few nanometers thick) with a reduced Young's

modulus of 50 GPa covered the work-hardened steel substrate (tests A and B on figure 6).
- A thicker layer (15 to 30 nm) with a reduced Young's modulus of 50 – 80 GPa was found in other places (tests C and D on figure 6), sometimes with accommodation pressure effect (threshold pressure $H_0 = 4.8$ GPa). Such layer behaves like the sulphide-oxide layer of the ZDTP tribofilm [16].
- Elsewhere, the structure of the tribofilm was more complex, with a soft layer covering a stiffer one. For example, test E on figure 6 corresponds to a soft layer, 12 nm thick, with properties comparable to those of the MoDTC tribofilm (hardness of 0.2 GPa and reduced Young's modulus of 5 GPa) which covers a stiffer layer, 18 nm thick, with a reduced Young's modulus of 50 GPa.

This heterogeneity was confirmed by the indentation tests conducted on the solvent-washed ZDTP + MoDTC tribofilm, where, at least, three different types of film were identified:
- In some places, the film behaved like a one layer system, able to accommodate the pressure (pressure threshold 2.8 GPa). Its thickness was between 35 nm and 150 nm. The surface hardness was about 2 – 3 GPa and the reduced Young's modulus was about 55 - 65 GPa.
- In other places, the film behaved like a bilayered structure: a surface layer, about 25 nm thick, with properties comparable to those of the MoDTC tribofilm (hardness of 0.3 - 0.4 GPa, reduced Young's modulus of 8 GPa), covers a stiffer layer, 150 nm thick, with a reduced Young's modulus of about 80 GPa.
- Elsewhere, the surface film was between 3 and 15 nm thick, with properties comparable to the lower properties measured on the ZDTP tribofilm (hardness about 1 – 1.5 GPa and reduced Young's modulus about 10 GPa). For some tests, this surface film was able to accommodate the pressure, with a pressure threshold of 1 – 1.5 GPa. It covers a stiffer layer, 10 to 55 nm thick, with a reduced Young's modulus varying from 60 to 110 GPa.

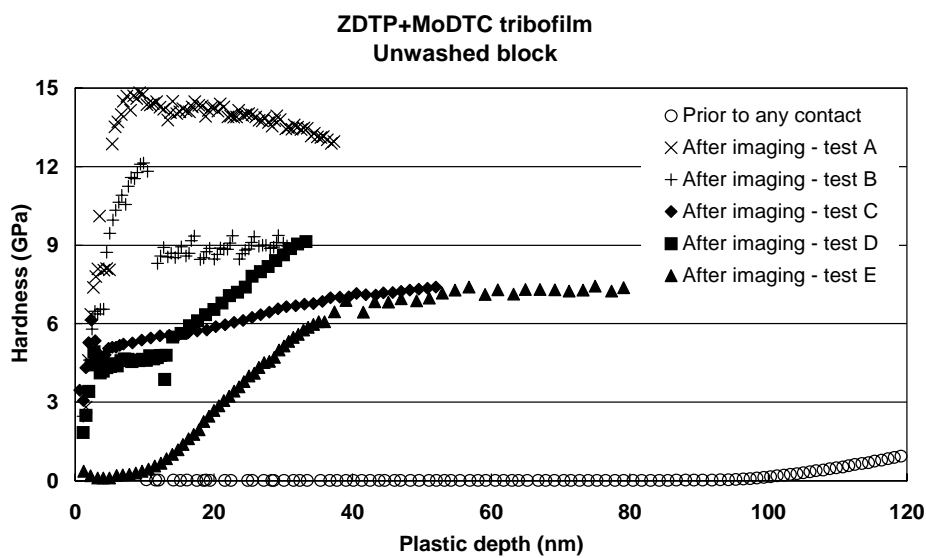

*Figure 6:* *Representative hardness curves obtained on the unwashed ZDTP + MoDTC tribofilm, prior to any contact and after the imaging procedure. The film is spatially heterogeneous in thickness and in mechanical properties.*

ZDTP + MoDTC + detergent/dispersant tribofilms ("full formulation" tribofilms)
Nanoindentation tests performed in fresh areas, prior to any contact showed that, at the surface of the unwashed tribofilm, there was a very soft layer, mobile under the diamond tip, with an apparent thickness of a few hundreds of nanometers.

Representative hardness curves obtained on the unwashed block near these initial contacts are shown figure 7. Contrary to the ZDTP + MoDTC tribofilm, the film was found to be spatially homogeneous. Only its thickness was found to vary, depending on the tested area. A very thin softer layer was detected at the surface of the tribofilm, which did not resist to imaging nor scanning, except if the normal load was very low (lower than 0.3 µN). This layer had a hardness value (about 0.3 – 0.4 GPa) comparable to the hardness value of the MoDTC tribofilm. The observed large hardness increase when the load increased also indicated that the tribofilm had a great capability to accommodate the applied pressure. This result was confirmed by the interpretation of the stiffness measurements using the rheological model, which also showed that the tribofilm had a complex structure. At its surface, there was first a layer with a thickness of only a few nanometers (2 nm to 7 nm) and a reduced Young's modulus of 10 - 15 GPa. Then, there was a second layer (thickness between 20 nm and 140 nm) with a higher reduced Young modulus of 65 – 80 GPa.

A similar tribofilm was tested after n-heptane washing. It also had a great ability to accommodate the applied pressure. From the stiffness measurements, on most places, the film was found to behave like a film constituted by two layers. The surface layer was thin (5 to 25 nm) with a reduced Young's modulus value in the range 15 – 20 GPa. The thickness of the underlayer was found to vary between 0 (no underlayer, example of figures 1 and 2) and 100 nanometers and its elastic modulus was in the range 110 - 120 GPa.

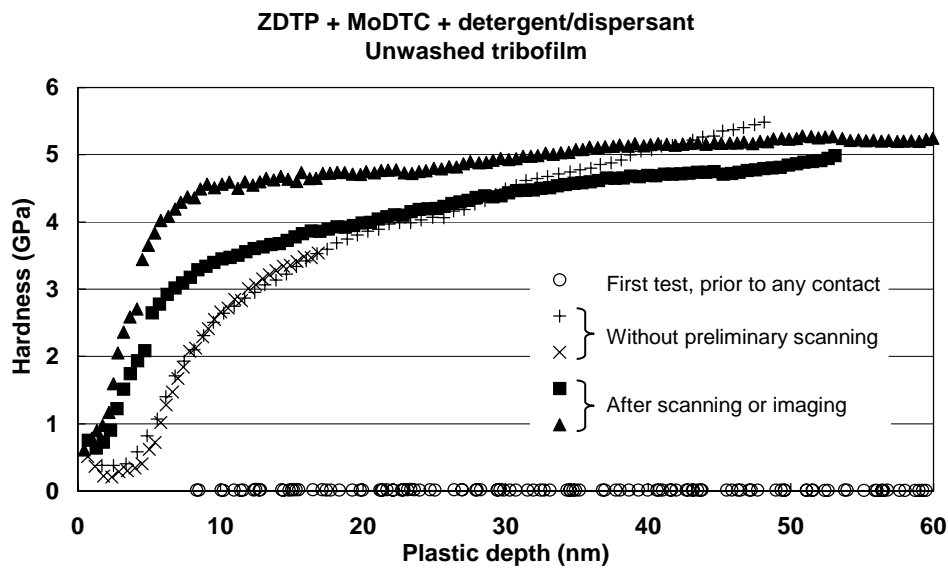

*Figure 7:* *Representative hardness curves obtained on the unwashed ZDTP + MoDTC + detergent/dispersant tribofilm ("full formulation"), prior to any contact and in the region near the first contacts, either without preliminary surface scanning or after scanning/imaging procedure.*

Figure 8 compares representative hardness curves for all tested tribofilms. For the ZDTP + MoDTC tribofilms, three curves are plotted because of the variety of obtained results revealing the spatial heterogeneity of this tribofilm. A representative hardness curve for the ZDTP tribofilm tested in the same conditions in a previous study [16] has been added for comparison.

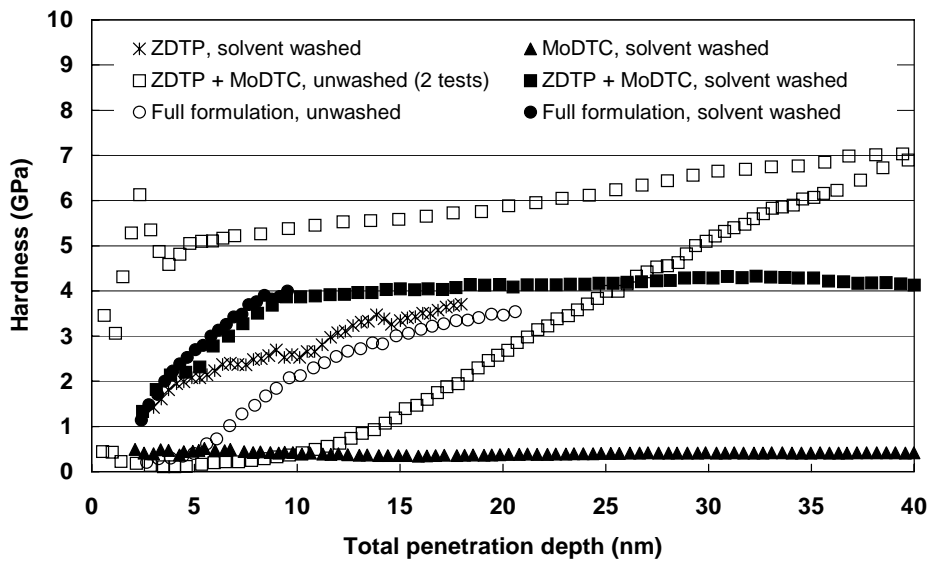

*Figure 8:* *Comparison of the hardness curves obtained on the different tribofilms. The hardness curve obtained for a ZDTP anti-wear tribofilm obtained from a previous study is plotted for comparison.*

*4.2. Nanofriction experiments*

Nanofriction experiments were conducted on the three preceding tribofilms and also on a ZDTP tribofilm and on a ZDTP + detergent/dispersant tribofilm. In order to simplify the following graphs, only one representative curve was plotted for each tribofilm (or two when it was necessary to illustrate the dispersion when it was significant).

Figure 9 shows the evolution of the friction force versus the normal force for the tested tribofilms. For a given formulation, there was very little difference between the results obtained on unwashed and on solvent washed tribofilms at low load, indicating that the solvent washing does not seem to affect the frictional behaviour of the tribofilm. This agrees with the idea that the soft viscous overlayer is supposed to serve as precursor for the tribofilm rather than that it plays a mechanical role during friction.

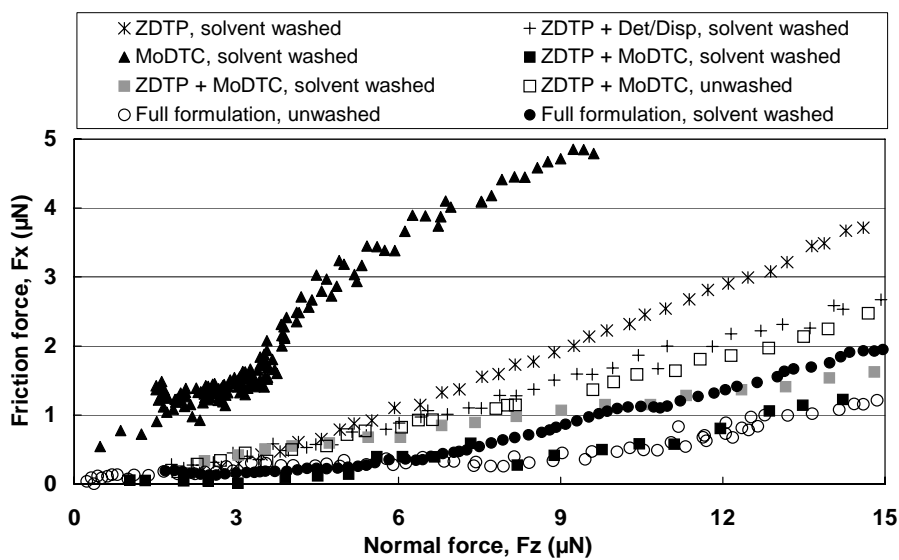

*Figure 9:* *Friction force (Fx) versus normal force (Fz) during nanofriction tests with increasing penetration depth for different tribofilms.*

It is also worth noting that the heterogeneity in mechanical properties found on the ZDTP + MoDTC tribofilm also exists in the frictional properties. For this tribofilm, the friction force at low normal loads may be comparable either to the friction force obtained for the ZDTP tribofilm or to the friction force obtained for the "full formulation" tribofilm.

Under the present testing conditions, it can be observed that the lower friction forces were obtained for films containing MoDTC together with ZDTP. The higher were obtained for the tribofilm from MoDTC alone.

Figure 10 shows the evolution of the friction coefficient versus mean pressure. The existence of low friction coefficient values ($0.01<\mu<0.05$) appears to be related both to the presence of MoDTC additive in the initial lubricant and to the ability for the tribofilm to reach sufficiently high pressure values (1.5 – 3 GPa) during the friction test. Thus, the MoDTC tribofilm, which is not able to resist to the contact pressure by increasing its mechanical properties seems to be ineffective in reducing friction, contrary to the tribofilms containing ZDTP and MoDTC together, which are able to accommodate the contact pressure by increasing their mechanical properties. Nevertheless, both behaviours (high or low friction) were observed for the ZDTP + MoDTC tribofilms. This is certainly due to the spatial heterogeneity of these tribofilms, which behave on some places like ZDTP tribofilms, or elsewhere like "full formulation" tribofilms. It was also observed that tribofilms formed without MoDTC were ineffective in reducing friction even if high contact pressures were reached during the friction tests.

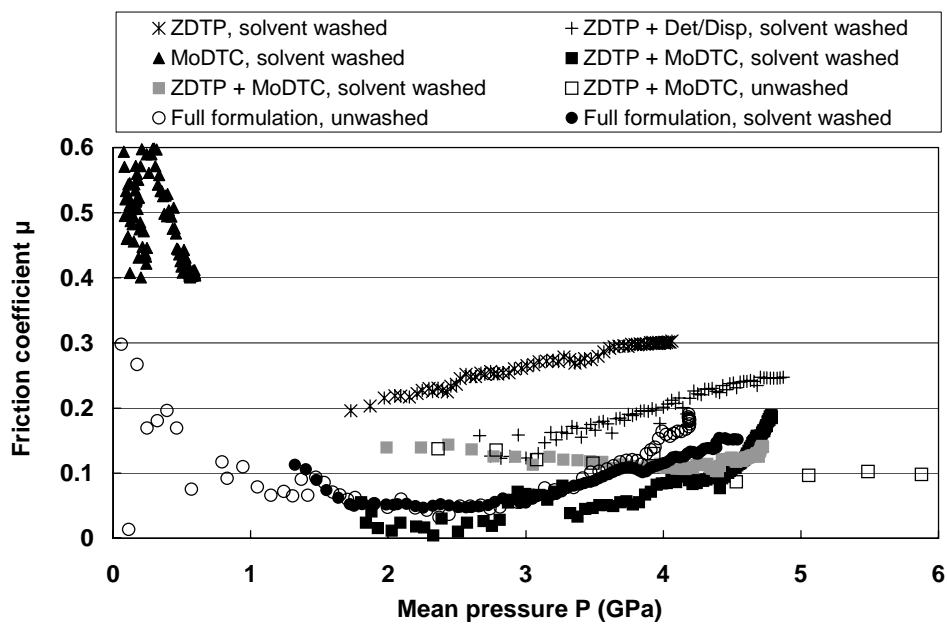

*Figure 10:* Apparent friction coefficient versus mean pressure for the different tested tribofilms.

When the evolution of the friction coefficient is plotted versus penetration depth (figure 11), it appears that, when it existed, the low friction coefficient domain was detected a few nanometers below the surface of the tribofilm. It also shows that, for the full formulation, the low friction domain was deeper for the unwashed tribofilm than for the solvent washed one. The unwashed tribofilm appears to be covered by a surface layer with rather bad frictional properties, which can be removed by solvent washing or by "mechanical" sweep (low load scanning procedures for example).

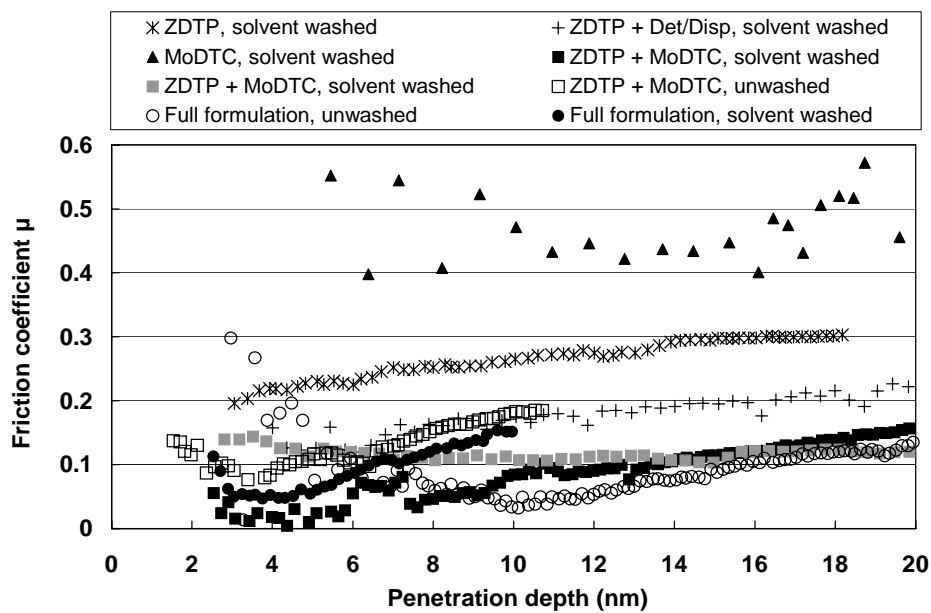

*Figure 11:* *Apparent friction coefficient versus penetration depth for the different tested tribofilms.*

## 5. Discussion

Because of the inhomogeneous and patchy nature of anti-wear tribofilms and of their low thickness, very few results are published concerning their mechanical properties [24-28]. Moreover, the differences in sample preparation and the diversity of used techniques and experimental procedures render delicate the comparison of the obtained results. For example, the Young's modulus values given by Aktary et al. for a ZDTP tribofilm [28] are significantly higher that those we measured but one explanation can be that they did not take into account the substrate's elasticity in their calculations, contrary to what is done in the current study. Or if we attempt to compare our results with those recently published by Ye et al. on ZDTP and ZDTP + MoDTC tribofilms [14, 15], this reveals significant differences. For example, Ye et al. found that both tribofilms possess the same hardness and modulus depth distributions, corresponding to continuously and functionally graded materials, when in the present work, the hardness curves for similar tribofilms did not coincide and the use of our rheological film model allowed us to describe the tribofilms as layered materials with properties adaptable to contact conditions. The hardness and modulus values, respectively 10 GPa at a contact depth of 30 nm and 215 GPa at a depth of 20 nm, that they reported are also significantly higher than those we measured and also higher than those given by Aktary et al. This could be due to differences in sample preparation and also certainly to the use of different methods and assumptions for the treatment of the nanoindentation data.

Concerning the frictional behaviour of the tribofilms, the presented nanofriction tests were conducted in unlubricated conditions, at very low speed (2 to 5 nm/s) and the measured nanofriction coefficients corresponded to the friction between the diamond tip and the tribofilm (over its steel substrate). That is why it also seems difficult to compare our values to macroscopic friction coefficient values obtained on classical tribometers. The latter are representative of steel on steel contact in the presence of a tribofilm and are averaged over the whole contact surface. However, our local values are not far from the end of test Amsler macroscopic friction coefficient values published by Pidduck and Smith [25] for ZDTP, ZDTP + detergent/dispersant and ZDTP + friction modifier tribofilms. Moreover, these macroscopic values were found to be proportional, with a factor 0.7, to micro-friction coefficient values measured with Lateral Force Microscopy by the same authors, making them suggest that there may be a link between macro and micro-frictional behaviour of smooth regions of anti-wear tribofilms. Unfortunately, no tribofilm obtained from friction modifier alone were tested in this study, with which we could compare our results. Nevertheless, macroscopic friction coefficient values, in the range 0.10 – 0.14, measured on an alternative ball on plane tribometer were reported by Muraki and Wada [6] for oil containing MoDTC alone. They conclude that such lubricant was ineffective in reducing friction, contrary to the oil containing MoDTC together with ZDTP. More recently, similar high macroscopic friction coefficient values (in the range 0.095 – 0.2) were measured by Unnikrishnan et al. for oil containing MoDTC alone [29]. On the other hand, Grossiord et al. reported very low steady-state friction coefficient (0.04) measured for base oil + MoDTC during SRV friction tests, and a lower steady-state value (0.02) for friction tests in a UHV tribometer, carried out by sliding a macroscopic hemispherical steel pin again a flat covered by a MoDTC tribofilm [13]. From tests carried out in a high frequency reciprocating rig, Graham et al. [30] also reported that, in the absence of ZDTP, MoTDC additives were effective in reducing friction at a combination of high additive concentration and high temperature (up to 0.4% wt. and 200°C). Such diversity of results, certainly partly due to the various tests conditions, makes unreasonable a comparison between the very high nanofriction coefficient measured on the MoDTC tribofilm under the present testing conditions and those published values. As, regarding the literature, the formation of $MoS_2$ was well established for MoDTC containing lubricants, the question is how can we explain such high friction coefficient during the nanofriction tests ? Or what caused the very low friction observed when ZDTP was used together with MoDTC ? From figure 10, the low friction coefficient values ($0.01<\mu<0.05$) were observed for the MoDTC containing lubricants when the contact pressure was in the range 1.5 – 3 GPa (the question of the spatial heterogeneity of the ZDTP + MoDTC tribofilm will be discussed latter). These high pressures were measured for tribofilms able to increase their mechanical properties, thus accommodating the contact conditions, which was demonstrated to be the case for ZDTP anti-wear tribofilms [16]. On the other hand, high pressures were not reached for the soft MoDTC tribofilm. Thus, the easy sliding of the $MoS_2$ sheets could result from a favourable orientation induced by sufficiently high contact pressure values. The ability of $MoS_2$ sheets to orient in a favourable direction was reported by Grossiord et al. [31] and Martin et al. [32], who recently investigated tribochemical interactions between ZDTP, MoDTC and OCB (overbased detergent calcium borate) additives. Using high resolution TEM observations of wear debris, coupled with wear scar micro-spot XPS analysis, they observed perfectly oriented $MoS_2$ sheets, with their basal plane parallel to the flaky wear fragments. Such "mechanical" interpretation of the role of the contact pressure agrees with previous work of Muraki et al. who studied the effect of roller hardness on the rolling sliding characteristics of MoDTC in the presence of ZDTP and concluded that the friction reduction effect increased with higher degree of roller hardness [10]. Yamamoto also reported that a necessary condition for improving the friction and wear characteristics of a lubricant was the formation of surface films composed of iron phosphates with high hardness and Mo-S compounds [11]. Concerning the spatial heterogeneity of the ZDTP + MoDTC tribofilms, it can be worth noting that using high resolution TEM observations of wear debris collected after friction tests, coupled with AES and XPS studies of rubbing surfaces, Grossiord et al. described the ZDTP + MoDTC tribofilm as being composed of a mixture of glassy zinc phosphate zones containing molybdenum, and carbon-rich zones containing zinc and highly-dispersed $MoS_2$ single sheets [13, 33].

The observation that, during the nanofriction tests, the low friction domain was located a few nanometers below the surface also corroborates this interpretation. As the nanofriction tests were conducted at increasing depth, the sufficiently high pressures were obtained after a few nanometers penetration depth inside the $MoS_2$ containing layer (with properties similar to the MoDTC tribofilm), thanks to the presence of the underneath resisting anti-wear layer, whose characteristics are similar to those of the phosphate layer of the ZDTP tribofilm.

Finally, combining the results obtained from the nanoindentation and nanofriction experiments, we can propose a possible schematic description of the anti-wear tribofilms obtained from the "full formulation" oil. Some assumptions are also made on what happened during nanofriction tests on such tribofilms (see figure 12 on which for convenient drawing, as the Berkovitch diamond tip is not sharp, it was represented by a flat punch).

A soft layer containing non-oriented $MoS_2$ sheets is present at the surface of the tribofilm (layer (a) in figure 12). This layer, 0 to 25 nm thick, has mechanical properties comparable with those of the MoDTC tribofilm (0.3 – 0.5 GPa for the hardness and 3 – 10 GPa for the reduced Young's modulus). Its friction coefficient is rather high. This layer is easily damaged or removed by the diamond tip during imaging or line-scanning procedures. When the contact pressure is sufficiently high, friction induces a favourable orientation of the $MoS_2$ sheets, over a thickness of 1 or 2 nanometers (layer (b) in figure 12), resulting in very low friction coefficient values which combine with the anti-wear efficiency of the tribofilm. Under this layer, there is then an anti-wear layer (layer (c) in figure 12), with properties similar to those of the polyphosphate layer of the ZDTP tribofilm. Then, just over the substrate (noted (e) in figure 12), there is a bonding layer (layer (d) in figure 12) with high mechanical properties (oxides, sulfides).

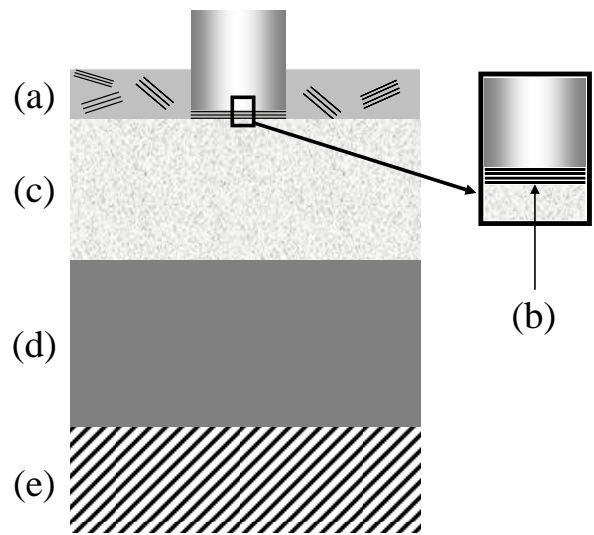

*Figure 12:* *Possible schematic description of the anti-wear tribofilm obtained from the "full formulation" and orientation of the $MoS_2$ planes of the outer layer resulting from a nanofriction tests (for convenient drawing, as the Berkovitch diamond tip is not sharp, it was represented by a flat punch). The thickness of each layer is arbitrary drawn as it varies significantly depending on the tested area (from zero when the layer is not present to a few tens of nanometers).*

(a) *Soft layer containing non-oriented $MoS_2$ sheets, with mechanical properties comparable to those of the MoDTC tribofilm,*
(b) *Layer of favourably frictionally oriented $MoS_2$ sheets with a typical thickness of 1 or 2 nm,*
(c) *Layer with properties similar to those of the polyphosphate layer of the ZDTP tribofilm,*
(d) *Bonding layer with high mechanical properties (oxides, sulfides),*
(e) *Steel substrate.*

## 6. Conclusions

Thanks to the combined used of (i) nanoindentation experiments with continuous stiffness measurements coupled with imaging procedures, (ii) a specifically developed rheological film model and (iii) nanofriction tests, synergistic effects of ZDTP and MoDTC on frictional behaviour of anti-wear tribofilms have been evidenced from mechanical considerations. One original feature of this study lies in the characterisation of unwashed anti-wear tribofilms with their full structure preserved.

The structure and nanomechanical properties (hardness and reduced Young's modulus) of tribofilms formed with different mixtures of additives (ZDTP, MoDTC, detergent/dispersant) were first determined.

Concerning the occurrence of very low friction ($0.01<\mu<0.05$), the contact pressure was found to be a critical parameter. The low friction coefficient values were attributed to a favourable orientation of $MoS_2$ sheets present in the outer layer of the tribofilms formed from MoDTC containing lubricants. Such a favourable orientation occurred only if sufficiently high contact pressure was reached. These high contact pressures were attained when ZDTP was used as oil additive together with MoDTC because one of the main characteristics of ZDTP additives is to form protective anti-wear tribofilms under boundary lubrication, with varying structure and properties with depth, among which is an amazing ability to increase their mechanical properties, thus accommodating the contact conditions.

A possible schematic description of the tribofilms containing both ZDTP and MoDTC was deduced and a mechanism was proposed to account for the mechanical synergy that occurs during nanofriction tests on such tribofilms.

**Aknowledgement**
The authors thank Shell Research Limited for financial support and permission to publish.